\RequirePackage{fix-cm}
\documentclass[smallextended]{svjour3}       
\smartqed  
\usepackage{amsmath,mathtools}
\usepackage{graphicx}
\usepackage{enumerate}
\usepackage{natbib}
\usepackage{hyperref}
\usepackage{url} 
\usepackage{multirow}
\usepackage[most]{tcolorbox}
\usepackage{float}
\usepackage[caption = false]{subfig}
 \journalname{Environmental and Ecological Statistics}

\numberwithin{equation}{section}
\numberwithin{lemma}{section}
\numberwithin{theorem}{section}
\numberwithin{proposition}{section}

\begin{document}

\title{Distribution-free changepoint detection tests based on the breaking of records
}


\author{Jorge Castillo-Mateo}


\institute{Jorge Castillo-Mateo \at
           Department of Statistical Methods, University of Zaragoza, Zaragoza, Spain \\
           ORCID: 0000-0003-3859-0248\\
           \email{jorgecm@unizar.es}
}

\date{Received: date / Accepted: date}

\maketitle

\begin{abstract}
The analysis of record-breaking events is of interest in fields such as climatology, hydrology or anthropology. In connection with the record occurrence, we propose three distribution-free statistics for the changepoint detection problem. They are CUSUM-type statistics based on the upper and/or lower record indicators observed in a series. Using a version of the functional central limit theorem, we show that the CUSUM-type statistics are asymptotically Kolmogorov distributed. The main results under the null hypothesis are based on series of independent and identically distributed random variables, but a statistic to deal with series with seasonal component and serial correlation is also proposed. A Monte Carlo study of size, power and changepoint estimate has been performed. Finally, the methods are illustrated by analyzing the time series of temperatures at Madrid, Spain. 

The R package \texttt{RecordTest} publicly available on CRAN implements the proposed methods. 

\end{abstract}


\section{Introduction}
\label{sec:intro}

An observation in a time series is called an upper (lower) record if it is greater (smaller) than all previous observations in the series. Therefore a new record is a remarkable event that attracts great attention in numerous applications, whether in environmental fields, economy, sports, physics or biology \citep[see, e.g.,][and references therein]{wergen2013}. Particularly interesting is the study of record events in environmental sciences and their connection with climate change. For example, \cite{benestad2004} compared the observed and expected number of records under stationarity by means of a $\chi^2$-test and graphical tools. Respectively, \cite{coumou2013} and \cite{lehmann2015} found an increase in temperature and precipitation record-breaking events with respect to a stationary climate on a global scale. In addition to its many applications, the main foundations in the framework of theory of records can be found in the monographs \cite{arnold1998} and \cite{nevzorov2001}.

An aspect of interest is the study of the evolution of the number of records over time, in particular the identification of changes in their behavior. To analyze this type of change, changepoint detection methods that make use of the record occurrence should be considered.

The changepoint problem tries to identify times when the probability distribution function of a time series changes. In general the problem concerns both detecting whether or not a change has occurred and identifying its time of occurrence. Although several changes might be considered, our work resides in the at most one changepoint (AMOC) domain. The first results on changepoint detection start with \cite{page1954,page1955} who introduced a cumulative sum (CUSUM) statistic to locate a shift in the mean of independent and identically distributed (IID) normal random variables (RVs). Since then, several methods have been proposed, many of which can be found in the monographs \cite{brodsky1993} and \cite{csorgo1997}. Noteworthy is the importance of changepoint detection techniques in climatology \citep{reeves2007}, but also in very different fields such as economy, speech processing, etc.

Traditional changepoint detection methods attempt to find changes in location or scale, more recently, changepoint detection in the extreme values has also been an active area of research. For example, \cite{dierckx2010} introduced tests to detect changes in the parameters of the generalized Pareto distribution based on its likelihood for models of excesses over threshold, \cite{kojadinovic2017} studied several tests for independent samples of block maxima, and \cite{silva2020} proposed a changepoint model for the r-largest order statistics. \cite{ratnasingam2021} proposed procedures based on the modified information criterion and the confidence distribution for detecting changepoints in the three-parameter Weibull distribution. Non-homogeneous Poisson processes have also been considered to study changepoints in the occurrence of peaks over threshold  \citep{achcar2010,achcar2016,rodrigues2019}. To the best of our knowledge, there is no changepoint detection method based on the breaking of records. There are, however, tests for trend detection based on the breaking of records. \cite{foster1954} proposed two simple statistics based on the number of records to test the hypothesis that $T$ observations have been independently drawn from the same continuous distribution. These tests were later improved by \cite{diersen1996} and more recently new tests and graphical tools were introduced by \cite{cebrian2021}.

The aim of this paper is to develop changepoint detection tests based on the record occurrence to detect changes in the tails of the distribution. The first use of the tests introduced in this paper is to detect changes in the record occurrence and therefore in the extreme values, however, they are also useful against other types of change such as a change in location or scale. When there is a gradual change in location or scale, it will generally take time to be significantly reflected in a change in the behavior of the number of records, so the second use of the proposed methodology lies in analyzing how long it takes a series from when a changepoint is detected using another method \citep[see, e.g.,][for a change in location]{pettitt1979}, until that change is reflected in the observed records. Beyond its theoretical and descriptive interest, the third use of these changepoint detection tests based on records is that they would be uniquely appropriate whenever the original data are not available while records are.

The proposed tests make use of CUSUM-type statistics based on the record indicator RVs. The functional central limit theorem for independent but nonidentically distributed RVs is used to show that the functional evolution of the number of records adequately standardized behaves asymptotically as a Wiener process and, as a consequence, the CUSUM-type statistics follow the Kolmogorov distribution. This characterization allows to obtain exact p-values for the tests. The use of weights in the statistics can improve the power of the tests under certain scenarios. However, we prove that the weighted statistics do not have the same asymptotic properties as the previous ones and the p-value must be calculated using Monte Carlo techniques. An approach to analyze series with seasonal component or serial correlation is also proposed. The statistics based on the record indicators will allow studying the extreme values of the distribution with the advantage of not needing the specification of an underlying distribution for the data, i.e., they are distribution-free. Also, the requirement on the variance of data as in other CUSUM-type statistics is avoided here.

The rest of the paper is organized as follows. Section~\ref{sec:theory} introduces our records statistics, establishes their asymptotic distribution under the null hypothesis and proposes some generalizations. Section~\ref{sec:simulation} compares these tests under various scenarios by means of Monte Carlo simulations. An application to temperature data is presented in Section~\ref{sec:application}, and Section~\ref{sec:conc} concludes the paper with final comments, conclusions and future work. 

Finally, note that the proposed tests for changepoint detection are available from the R \citep{R} package \texttt{RecordTest} \citep{recordtest}.

\section{Tests based on theory of records}
\label{sec:theory}

Let $X_1, \ldots, X_T$ be a sequence of IID continuous RVs. The sequences of upper and lower record indicators, $(I_t)$ and $(I_t^L)$, are defined by $I_1 = I_1^L = 1$ and for $t=2,\ldots,T$, by
\begin{equation*}
I_t =
\begin{cases} 
1 & \text{if } X_t > \max\{X_1, \ldots, X_{t-1}\}, \\
0 & \text{otherwise}, 
\end{cases}
\quad
I_t^L =
\begin{cases} 
1 & \text{if } X_t < \min\{X_1, \ldots, X_{t-1}\}, \\
0 & \text{otherwise}. 
\end{cases}
\end{equation*}
The sequence of differences in the upper and lower record indicators, $(d_t)$, is given by $d_t = I_t - I_t^L$, while the sequence of sums, $(s_t)$, is given by $s_t = I_t + I_t^L$.

The following lemma is a well known distribution-free result within the theory of records that characterizes the distribution of the record indicators, equally valid for upper and lower records \citep{arnold1998,nevzorov2001}.
\begin{lemma}
	Let $X_1, \ldots, X_T$ be a sequence of IID continuous RVs. Then, the record indicators $I_1, \ldots, I_T$ are independent and
	\begin{equation*}
	p_t = P(I_t = 1) = \frac{1}{t}, \quad t = 1,\ldots,T.
	\end{equation*}
\end{lemma}
It is easily checked that the expectations and variances for $t = 2,\ldots,T$, are 
\begin{align*}
E(I_t) &= \frac{1}{t}, \quad 
Var(I_t) = \frac{1}{t} \left( 1-\frac{1}{t} \right), \\ 
E(d_t) &= 0, \quad
Var(d_t) = \frac{2}{t}, \\
E(s_t) &= \frac{2}{t}, \quad 
Var(s_t) = \frac{2}{t} \left( 1-\frac{2}{t} \right).
\end{align*}

Given $p_t$ the probability of upper or lower record at time $t$, our aim is to construct asymptotic tests with null hypothesis
\begin{equation*}\label{eq:H0}
\mathcal{H}_0: p_t = 1 / t, \quad 1 \le t \le T,
\end{equation*}
against the two-sided alternative hypothesis given by
\begin{equation}\label{eq:H1}
\mathcal{H}_1: p_t = 1 / t, \quad 1 \le t \le t_0 \quad \text{and} \quad p_t \neq 1 / t, \quad t_0 < t \le T,
\end{equation}
where $t_0$ denotes the time of a possible change in the probabilities of observing new records with respect to the stationary case. The alternative hypothesis supports many nonstationary scenarios, for example a shift or a drift in location, variation or in one or both tails.

\subsection{Tests based on asymptotic results}
\label{sec:asymptotic}

To obtain a p-value from a changepoint detection test, the exact distribution of the changepoint statistic is usually impractical, so it is generally preferable to have asymptotic results. Wiener processes, Brownian bridges and other Gaussian processes arise as asymptotic distributions in many limit problems providing exact tail probabilities. Our first objective is to build from the indicators above a random function $W_T(\nu)$, for $\nu \in [0,1]$, in such a way that $W_T(\nu)$ converges in distribution to a Wiener process. For this purpose, we define the standardized record indicators, $\xi_{T1},\ldots,\xi_{TT}$ as
\begin{equation}\label{eq:xi}
\xi_{Tt} = \frac{I_t - E(I_t)}{\sigma_T},
\end{equation}
where $\sigma_t^{2} = \sum_{k=1}^{t} Var(I_k)$. We also define the standardized number of records $S_{Tt} = \sum_{k=1}^{t} \xi_{Tk}$, its variance $\nu_{Tt} = \sum_{k=1}^{t} Var(\xi_{Tk}) = \sigma_t^2 / \sigma_T^2$, and finally the random function
\begin{equation}\label{eq:process}
W_T(\nu) = 
S_{Tt} + \xi_{T,t+1} \frac{\nu - \nu_{Tt}}{\nu_{T,t+1} - \nu_{Tt}}
\end{equation}
for $\nu \in [\nu_{Tt}, \nu_{T,t+1}]$. Note that $S_{T1}=0$, $\nu_{T1}=0$ and $\nu_{TT}=1$. It is noteworthy that the function $W_T(\nu)$ is a random broken line connecting points in the plane with coordinates $(\nu_{Tt},S_{Tt})$ for $t=1,\ldots,T$.

One of our major results is the asymptotic characterization of the functional evolution of the standardized number of records, $W_T(\nu)$, as a Wiener process. The result is essentially a consequence of the functional central limit theorem for independent but nonidentically distributed RVs \citep[see, e.g.,][]{gikhman1969}. To be under the conditions of the theorem, Lindeberg's condition needs to be proved for the variables $\xi_{Tt}$ in \eqref{eq:xi}, which follows immediately from
\begin{equation*}\label{eq:dem}
\lim_{T \rightarrow \infty} \sum_{t=1}^T E\left( \xi_{Tt}^2 \times \textbf{1}_{\{|\xi_{Tt}| > \varepsilon\}} \right) \le \lim_{T \rightarrow \infty} \textbf{1}_{\{1/\sigma_T > \varepsilon\}} = 0
\end{equation*}
for all $\varepsilon > 0$, where $\textbf{1}_{\{\cdot\}}$ is the indicator function.
\begin{theorem}\label{theoremW}
	Let $X_1,\ldots,X_T$ be a sequence of IID continuous RVs with $W_T(\nu)$ in \eqref{eq:process}. Then, as $T \rightarrow \infty$,
	\begin{equation*}
	W_T(\nu) 	\overset{\mathcal{D}} \longrightarrow W(\nu), \quad \nu \in [0,1],
	\end{equation*}
	in the metric space $\mathcal{C}[0,1]$, where $W(\nu)$ is a standard Wiener process.
\end{theorem}

Thus, the changepoint records statistic proposed is
\begin{equation}\label{eq:B*}
K_T = \max_{1\le t \le T} |B_T(\nu_{Tt})|,
\end{equation}
where $B_T(\nu) = W_T(\nu) - \nu W_T(1)$, $\nu \in [0,1]$. The time $t$ where \eqref{eq:B*} takes its maximum is the changepoint estimate $\hat{t}_0$. As a consequence of Theorem~\ref{theoremW}, $B_T(\nu)$ is asymptotically distributed as a standard Brownian bridge process. Moreover, the distribution of the supremum of the absolute value of a Brownian bridge is known as the Kolmogorov distribution. As $\sup_{0 \le \nu \le 1} |f(\nu) - \nu f(1)|$ is a continuous functional for $f$ in $\mathcal{C}[0,1]$, the asymptotic characterization under the null hypothesis of the statistic $K_T$ is as follows.
\begin{theorem}\label{theoremB}
	Let $X_1,\ldots,X_T$ be a sequence of IID continuous RVs with $K_T$ in \eqref{eq:B*}. Then, as $T \rightarrow \infty$,
	\begin{equation*}
	K_T
	\overset{\mathcal{D}} \longrightarrow
	K = \sup_{0 \le \nu \le 1} |B(\nu)|, 
	\end{equation*}
	where $B(\nu)$ is a standard Brownian bridge process and $K$ is a Kolmogorov distributed RV.
\end{theorem}

The null hypothesis is rejected when $K_T$ is too large to be explained by chance variation. In particular, if the alternative hypothesis in \eqref{eq:H1} is true for some time $t_0$, then it follows that $|B_T(\nu_{Tt_0})|$ is large and can show statistical evidence that a change occurred at time $t_0$. Under the null hypothesis, the p-value of the two-sided test can be calculated from any of the expressions of the Kolmogorov distribution
\begin{align*}
\label{eq:kolmogorov}
P(K \geq x) &= 
2 \sum_{k=1}^{\infty} (-1)^{k-1} \exp\left\{-2(kx)^2\right\}\\ &= 
1 - \frac{\sqrt{2 \pi}}{x} \sum_{k=1}^{\infty} \exp\left\{-\left(\frac{(2k-1) \pi}{2\sqrt{2} x}\right)^2\right\}.
\end{align*}

To give a clear interpretation of $K_T$, we define $N_t = I_1 + \ldots + I_t$ the number of records up to time $t$ and $N_{t_1:t_2} = I_{t_1} + \ldots + I_{t_2}$ the number of records between times $t_1 \le t_2$. Then, $B_T(\nu_{Tt})$ can be rewritten as
\begin{equation*}
B_T(\nu_{Tt}) = \frac{1}{\sqrt{Var(N_T)}} \left( (N_t - E(N_t)) - \frac{Var(N_t)}{Var(N_T)} (N_T - E(N_T)) \right).
\end{equation*}
Weighting for differences in the \emph{effective} sample sizes of the number of records in two segments, $\{1,\ldots,t\}$ and $\{t+1,\ldots,T\}$, $B_T(\nu_{Tt})$ can be viewed as a scaled difference between $Var(N_t)^{-1} (N_t - E(N_t))$ and $Var(N_{(t+1):T})^{-1} (N_{(t+1):T} - E(N_{(t+1):T}))$. Consequently, $K_T$ compares the number of records in both segments for every $t$ and assigns as estimator, $\hat{t}_0$, the point that separates the segment that deviates the most from the null hypothesis. The mean is $E(B_T(\nu_{Tt})) = 0$ and simple calculation leads to $Var(B_T(\nu_{Tt})) = \nu_{Tt} (1 - \nu_{Tt})$. The nonuniform variance, small when it is near the ends of $\{1,\ldots,T\}$, makes changepoints occurring near the beginning or the end of the series more difficult to detect (see Appendix~\ref{app:var} for further details). This is a common fact in CUSUM-type statistics.

The proposed statistic only uses the information from one tail of the distribution, the right tail if upper records are used or the left tail if lower records are used. To study both tails and collect more evidence with a single statistic, it is enough to consider the variables $d_t$ and $s_t$. Since the $d_t$'s and $s_t$'s also fulfill Lindeberg's condition, all the previous results are equally valid substituting $\xi_{Tt}$ in \eqref{eq:xi} by $\xi_{Tt} = d_t / \sigma_T$ with $\sigma_t^2 = \sum_{k=1}^{t} Var(d_k)$, $t=1,\ldots,T$; or respectively, $\xi_{Tt} = (s_t - E(s_t)) / \sigma_T$ with $\sigma_t^2 = \sum_{k=1}^{t} Var(s_k)$, $t=1,\ldots,T$. The statistic \eqref{eq:B*} based on $d_t$ can be used when an increase in upper records and a decrease in lower records are expected with respect to the null hypothesis, while the statistic based on $s_t$ can be used when an increase in both types of records is expected. In particular, the statistic based on $d_t$ can be useful against the alternative hypothesis of a trend in location, while the statistic based on $s_t$ can be useful against a trend in variation.

\subsection{Tests with weighted statistics} \label{sec:w}

Under the null hypothesis, the probability of record decreases as the series evolves. To give more importance to the most recent records and thus to be able to increase the power of the tests, we propose to give increasing weights, $\omega_t$, to the different records according to their position in the series as
\begin{equation} \label{eq:xiw}
\xi_{Tt}^{\omega} = \omega_t \frac{ I_t -E(I_t)}{\sigma_T},
\end{equation}
where $\sigma_t^2 = \sum_{k=1}^t \omega_k^2 Var(I_k)$, $t = 1,\ldots,T$. According to Proposition~\ref{prop:normal} (proved in Appendix~\ref{app:proof}), these variables do not in general have an asymptotically normal sum, so asymptotic results such as those of Theorem~\ref{theoremW} are not available. 
\begin{proposition} \label{prop:normal}
	Let $X_1,\ldots,X_T$ be a sequence of IID continuous RVs with the sequence of RVs $\xi_{Tt}^\omega$ in \eqref{eq:xiw} and $\omega_t \sim t^n$ as $t \rightarrow \infty$. If $n > 0$, then the central limit theorem does not hold for the $\xi_{Tt}^\omega$'s.
\end{proposition}
Likewise, a $K_T$-type statistic in \eqref{eq:B*} associated with the weighted variables can be defined, and the distribution of which can be simulated by means of Monte Carlo techniques under the null hypothesis. 

In this work we consider two different weights. First, linear weights $\omega_t = t-1$ \citep[see][for a detailed explanation]{diersen1996}. Second, weights that make the discrete sequence of times of the process, $\nu_{Tt} = \sigma_t^2 / \sigma_T^2$, $t=1,\ldots,T$, equally spaced, i.e., weights proportional to the inverse of the standard deviation (SD) of $I_t$, i.e., $\omega_1 = 0$ and $\omega_t = Var(I_t)^{-1/2} = t / \sqrt{t-1}$ for $t=2,\ldots,T$. These weights make the variance of $B_T(\nu_{Tt})$ symmetric in $\{1,\ldots,T\}$ (see Appendix~\ref{app:var}).

As above, the statistic has been defined in terms of the $I_t$'s but it is equivalent for the $d_t$'s or $s_t$'s. The SD in these cases suggests that the weights making the observed times of the process equally spaced are proportional to $\omega_1 = 0$, $\omega_t = \sqrt{t}$ for $t=2,\ldots,T$, for the statistic based on $d_t$; and $\omega_1 = \omega_2 = 0$, $\omega_t = t / \sqrt{t-2}$ for $t=3,\ldots,T$, for the statistic based on $s_t$.

\subsection{Tests for seasonal series}
\label{sec:seasonal}

\cite{hirsch1982} introduced a seasonal version for tests of randomness based on ranks. Following their ideas, we propose tests which are insensitive to the existence of seasonality and serial correlation. If the time series data of interest are daily (or monthly) data, then the null hypothesis of randomness where all the observations come from the same continuous distribution may be too restrictive. For example, most series of daily temperature or precipitation show very strongly the presence of seasonality and serial correlation. Let $\textbf{X} = (\textbf{X}_1,\ldots,\textbf{X}_M)$ be a sequence of series where $\textbf{X}_m = (X_{1m}, \ldots, X_{Tm})'$ is a series of RVs. That is, $\textbf{X}$ is the entire series, made up of subseries $\textbf{X}_1$ through $\textbf{X}_M$ (one for each day), and each subseries $\textbf{X}_m$ contains annual values from day $m$, for $m=1,\ldots,M$. Note that for further development the $M$ subseries must be independent, so in general a subset of these subseries will be used. That is, below a subset of independent subseries is considered, but the notation is maintained for simplicity. Then, we define the $t$-th upper record indicator for the $m$-th subseries as $I_{tm} = 1$ if $X_{tm} > \max\{X_{1m},\ldots,X_{t-1,m}\}$ and $I_{tm} = 0$ otherwise; analogously for lower records. That is, records are calculated independently for each subseries, and the null hypothesis is relaxed allowing observations of different subseries not to come from the same distribution. To define a $K_T$-type statistic that joins the information of all the subseries, we simply take the $\xi_{Tt}$'s in \eqref{eq:xi} as
\begin{equation*}
\xi_{Tt}^{\omega} = \omega_t \frac{\frac{1}{M}\sum_{m=1}^{M}  I_{tm} -  E(I_{t})}{\sigma_T},
\end{equation*}
where $\sigma_t^2 = \sum_{k=1}^{t} \omega_k^2 Var(I_k) / M$; or their respective versions based on $d_t$ or $s_t$. Thus, the alternative hypothesis is that of \eqref{eq:H1} with common changepoint $t_0$ for all the subseries. This approach not only allows the analysis of series with seasonal component, it also joins the information from several series, so the number of records and therefore the information used by the tests is greater.

\section{Monte Carlo experiments}
\label{sec:simulation}

We investigate the empirical size, power and changepoint estimate of the changepoint tests based on the records statistics introduced in Section~\ref{sec:theory}. Nine records statistics are considered: $N \equiv K_T$ in \eqref{eq:B*} with $\xi_{Tt}$ in \eqref{eq:xi}, $d$ and $s \equiv K_T$ in \eqref{eq:B*} substituting $I_t$ in \eqref{eq:xi} by $d_t$ and $s_t$, respectively; and the previous statistics with weights proportional to the inverse of the SD of $I_t$, $d_t$ and $s_t$, respectively (superscript \emph{var}); and linear weights $t-1$ (superscript \emph{linear}). Thus, three types of records statistics are analyzed. We denote by $N$-type statistics to the statistic $N$ and its weighted versions, equivalently for $d$ and $s$. Recall that, under the null hypothesis, the statistics $N$, $d$ and $s$ are asymptotically Kolmogorov distributed, while weighted statistics need Monte Carlo simulations to estimate their distribution ($1,000$ replicates are considered).

\subsection{Analysis of size}

We simulate $10,000$ replicates of $M$ independent series formed by $T$ independent samples from the standard normal distribution, i.e., 
\begin{equation*}
Y_{tm} = \epsilon_{tm} \sim N(0,1),
\quad \text{for } t = 1,\ldots,T \text{ and } m = 1,\ldots,M.
\end{equation*}
The size results are generalizable to any other continuous distribution given the distribution-free property of the tests under the null hypothesis. The size of the tests is simulated for the combination of values $T = 50,100$, $M = 1,12,36$ and for a large series $T=500$, $M=1$. 

Table~\ref{tab:size} reports the empirical size results of the changepoint tests based on the records statistics $N$, $d$ and $s$ for nominal values $\alpha=0.01,0.05,0.10$; i.e., we count how often the records statistics exceed the $99,95,90$-th percentile of the Kolmogorov distribution. We do not show the rejection frequencies of the tests based on the weighted statistics since their size is assured by simulating their p-value under the null hypothesis. All tests show an acceptable size for the levels $\alpha$ considered. Most of the tests are conservative, but their size approaches the nominal values as $T$ increases. When $M$ is greater than $1$, the size of the statistics is considerably less than the nominal value. The size of $d$ is particularly low, implying that these tests are very conservative.

In conservative tests, \cite{fisher2019} proposed a general method to obtain a size closer to the nominal value and therefore increase the power of the tests. For our proposed tests, the method simply consists of changing the $K_T$-type statistic by $-\sqrt{T} \log (1 - K_T/\sqrt{T})$. Although we do not apply this method in the present paper, it may be a factor to consider in applications with low evidence since the power can increase while maintaining a proper size.

\begin{table}[t]
	\begin{center}
		\caption{Test size for $\alpha=0.01,0.05,0.10$ level tests  \label{tab:size}}
		\begin{tabular}{ccccccccc}
			\hline\noalign{\smallskip}
			Statistic&$\alpha$&\multicolumn{7}{c}{$T|M$}\\
			&& $50|1$ &  $100|1$ &  $500|1$ & $50|12$ &  $100|12$ &  $50|36$ &  $100|36$  \\ 
			\noalign{\smallskip}\hline\noalign{\smallskip}
			\multirow{3}{*}{$N$}
			&$0.01$&$0.011$&$0.011$&$0.012$&$0.006$&$0.008$&$0.005$&$0.005$\\
			&$0.05$&$0.040$&$0.043$&$0.048$&$0.029$&$0.033$&$0.030$&$0.034$\\
			&$0.10$&$0.068$&$0.076$&$0.083$&$0.059$&$0.068$&$0.068$&$0.075$\\
			\noalign{\smallskip}\hline\noalign{\smallskip}
			\multirow{3}{*}{$d$}
			&$0.01$&$0.004$&$0.005$&$0.008$&$0.005$&$0.006$&$0.005$&$0.006$\\
			&$0.05$&$0.023$&$0.027$&$0.033$&$0.027$&$0.033$&$0.029$&$0.034$\\
			&$0.10$&$0.051$&$0.057$&$0.065$&$0.057$&$0.064$&$0.059$&$0.066$\\
			\noalign{\smallskip}\hline\noalign{\smallskip}
			\multirow{3}{*}{$s$}
			&$0.01$&$0.009$&$0.009$&$0.010$&$0.005$&$0.006$&$0.004$&$0.006$\\
			&$0.05$&$0.036$&$0.037$&$0.042$&$0.032$&$0.032$&$0.030$&$0.032$\\
			&$0.10$&$0.067$&$0.076$&$0.082$&$0.065$&$0.074$&$0.061$&$0.068$\\
			\noalign{\smallskip}\hline
		\end{tabular}
	\end{center}
\end{table}

\subsection{Analysis of power}
\label{sec:power}

The power analysis consists of $10,000$ simulations of $M$ independent series with $T$ observations following two scenarios under the alternative hypothesis.

\begin{description}
	\item[\textit{Scenario~A}.] Linear drift model in the mean:
	\begin{equation*}
	Y_{tm} = \mu_t + \epsilon_{tm}, \quad \text{for } t = 1,\ldots,T \text{ and } m = 1,\ldots,M,
	\end{equation*}
	where $\epsilon_{tm} \sim N(0,1)$, and $\mu_t = 0$ if $1 \le t \le t_0$ and $\mu_t = \theta (t - t_0)$ if $t_0 < t \le T$.
	\item[\textit{Scenario~B}.] Linear drift model in the SD: 
	\begin{equation*}
	Y_{tm} = \sigma_t \epsilon_{tm}, \quad \text{for } t = 1,\ldots,T \text{ and } m = 1,\ldots,M,
	\end{equation*}
	where $\epsilon_{tm} \sim N(0,1)$, and $\sigma_t = 1$ if $1 \le t \le t_0$ and $\sigma_t = 1 + \theta (t - t_0)$ if $t_0 < t \le T$.
\end{description}

We report results for $T=100$, $M = 1,12,36$, $t_0 = 25,50,75$ and the drift term $\theta = -0.10,-0.09,\ldots,-0.02,-0.01,-0.005,0.005,0.01,0.02,\ldots,0.09,0.10$ for Scenario~A and $\theta = 0.005,0.01,0.02,\ldots,0.09,0.10$ for Scenario~B. $N$-type statistics are analyzed against both scenarios, $d$-type statistics against Scenario~A and $s$-type statistics against Scenario~B.

\begin{figure}
	\begin{center}
		\includegraphics[width=\textwidth]{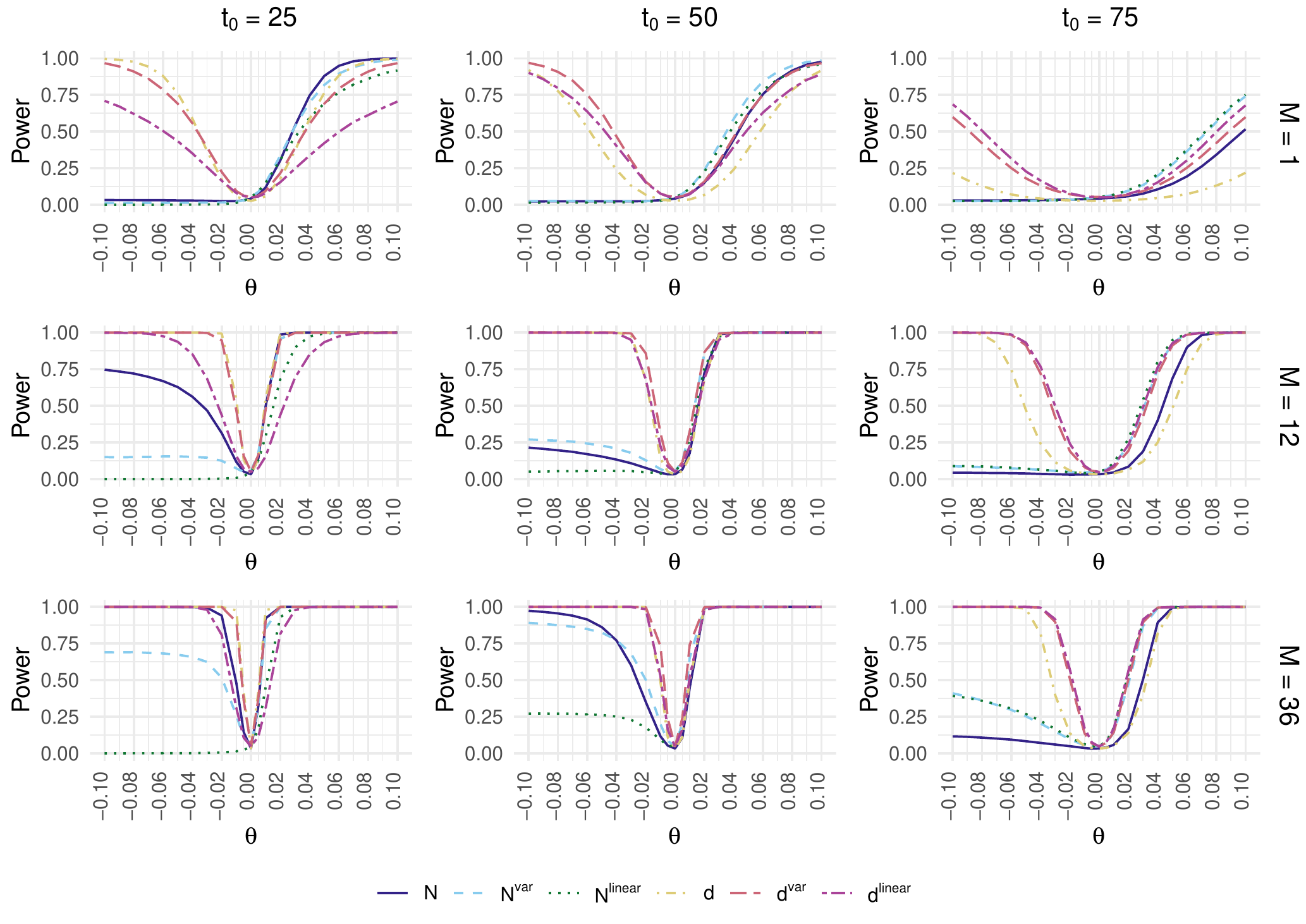}
	\end{center}
	\caption{Power functions of $N$ and $d$-type statistics for Scenario~A ($T = 100$) \label{fig:powerdrift_mean}}
\end{figure}

\begin{figure}
	\begin{center}
		\includegraphics[width=\textwidth]{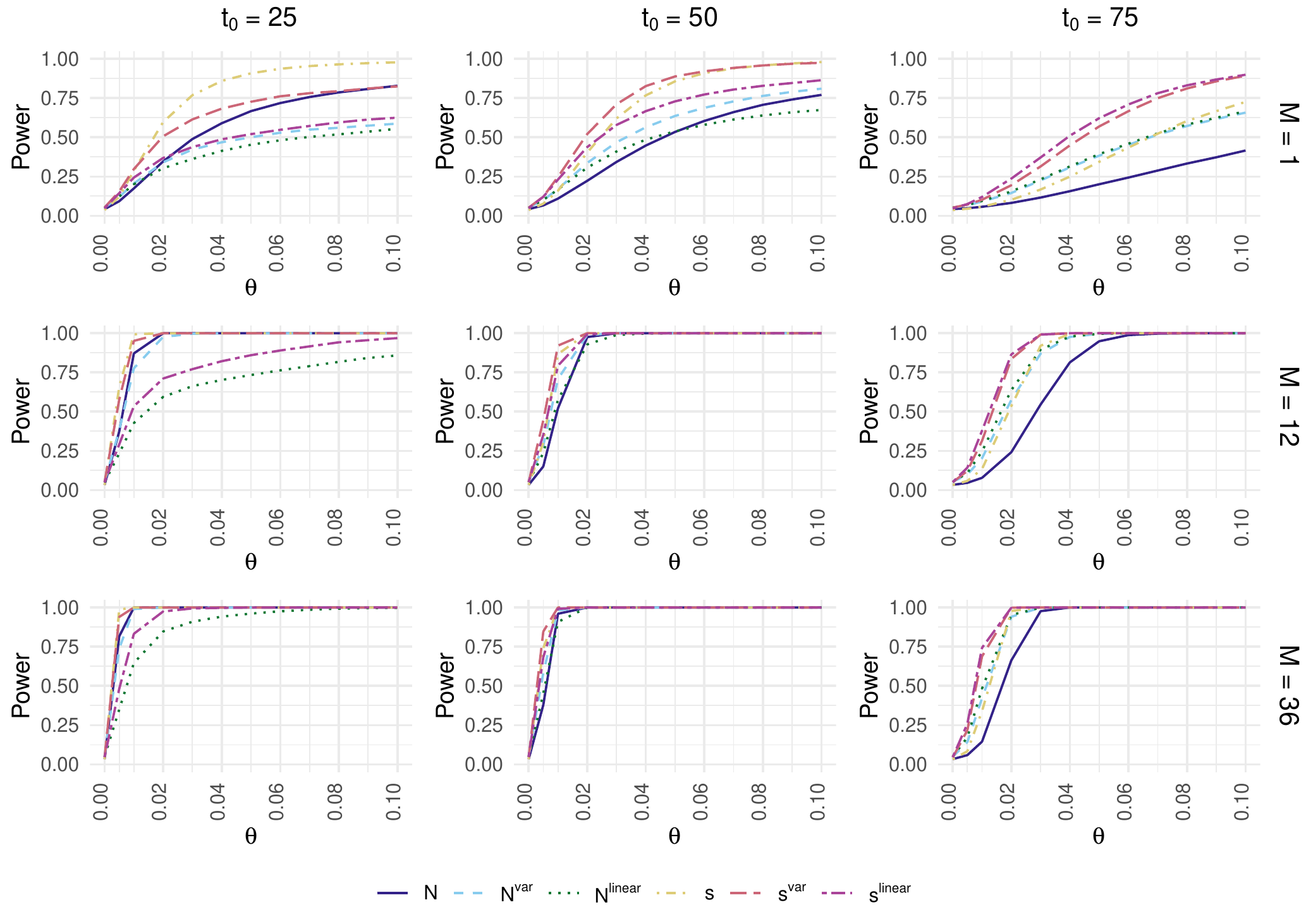}
	\end{center}
	\caption{Power functions of $N$ and $s$-type statistics for Scenario~B ($T = 100$) \label{fig:powerdrift_sd}}
\end{figure}

Figs.~\ref{fig:powerdrift_mean}~and~\ref{fig:powerdrift_sd} show, for $\alpha = 0.05$, plots of the power of the tests versus the trend $\theta$ for the Scenarios~A and B, respectively. We make the following observations:
\begin{enumerate}
	\item All tests increase their power as the magnitude of the drift $\theta$ or the number of series $M$ increases. In Scenario~A, $d$-type statistics have a symmetric behavior with respect to a vertical line at $\theta=0$, but when the drift is negative, $N$-type statistics have a power close to the nominal value unless $M$ is large. This phenomenon is due to the fact that the greatest effect that a negative trend can cause is that only one record is observed in each series and under the null hypothesis it is likely to find a single record in a small number of series but it is unlikely to find a single record in many series. Finally, note that the power of tests with upper records against a positive drift is equivalent to that of tests with lower records against a negative drift.
	
	\item The power of the statistics according to the position of the changepoint depends on the type of weight used. The tests have a lower power for a changepoint $t_0$ close to the end of the series, since the accumulated trend is smaller. The unweighted statistics have a higher power when the changepoint is at the beginning of the series and lose power as it approaches the middle and especially the end of the series. The statistics with weights proportional to the inverse of the SD, in Scenario~A, maintain the same power when the changepoint is in the first half of the series and lose power if the changepoint is at the end; in Scenario~B, they have a higher power when the changepoint is in the middle of the series. The statistics with linear weights have a higher power when the changepoint is in the middle or the end of the series than at the beginning. 
	
	\item For positive drifts and comparing statistics with the same type of weight. In Scenario~A, $N$-type statistics have a higher power than $d$-type for low $M$, but when $M$ is large this difference decreases and $d$-type have an equal or higher power than $N$-type. In Scenario~B, $s$-type statistics have a higher power than $N$-type.
	
	\item The statistics with weights proportional to the inverse of the SD turn out to have the overall best performance with the most balanced behavior. The statistics without weights are those that have a higher power when the changepoint is at the beginning of the series, the statistics with weights proportional to the inverse of the SD have a higher power when the changepoint is in the middle of the series and the statistics with linear weights have a higher power when the changepoint is at the end of the series. While the second show a power close to the best in each case, the first and third have considerably less power than the others when they are not the most powerful.
	
	\item Some cases in which the records tests reach a power between $0.85$ and $1$ for $T=100$ are given below. Under Scenario~A, when $t_0=25$, we would detect $\theta=0.05$ with $M=1$ for statistic $N$ or $\theta=0.02$ with $M=12$ for all statistics except those with linear weights or $\theta=0.01$ with $M=36$ for $d$. When $t_0=50$, we would detect $\theta=0.10$ with $M=1$ for all statistics or $\theta=0.03$ with $M=12$. Under Scenario~B, when $t_0=25$, we would detect $\theta=0.04$ with $M=1$ for the statistic $s$ or $\theta=0.01$ with $M=12$ or $\theta=0.005$ with $M=36$ for the statistics $s$ and $s^{var}$. When $t_0=50$, we would detect $\theta=0.05$ with $M=1$ or $\theta=0.01$ with $M=12$ for the statistics $s$ and $s^{var}$.
\end{enumerate}

\subsection{Analysis of changepoint estimation}

The analysis of the changepoint estimation reports results for Scenarios A and B considered in Section~\ref{sec:power} for $T=100$, $M=1$ and $\theta = 0.10$, and for $T=100$, $M=36$ and $\theta = 0.05$, both for a wide range of changepoints $t_0 = 10,20,\ldots,80,90$.

\begin{figure}
	\centering\captionsetup[subfloat]{labelfont=bf}
	\subfloat[$M=1$, $\theta = 0.10$]{\includegraphics[width=0.8\textwidth]{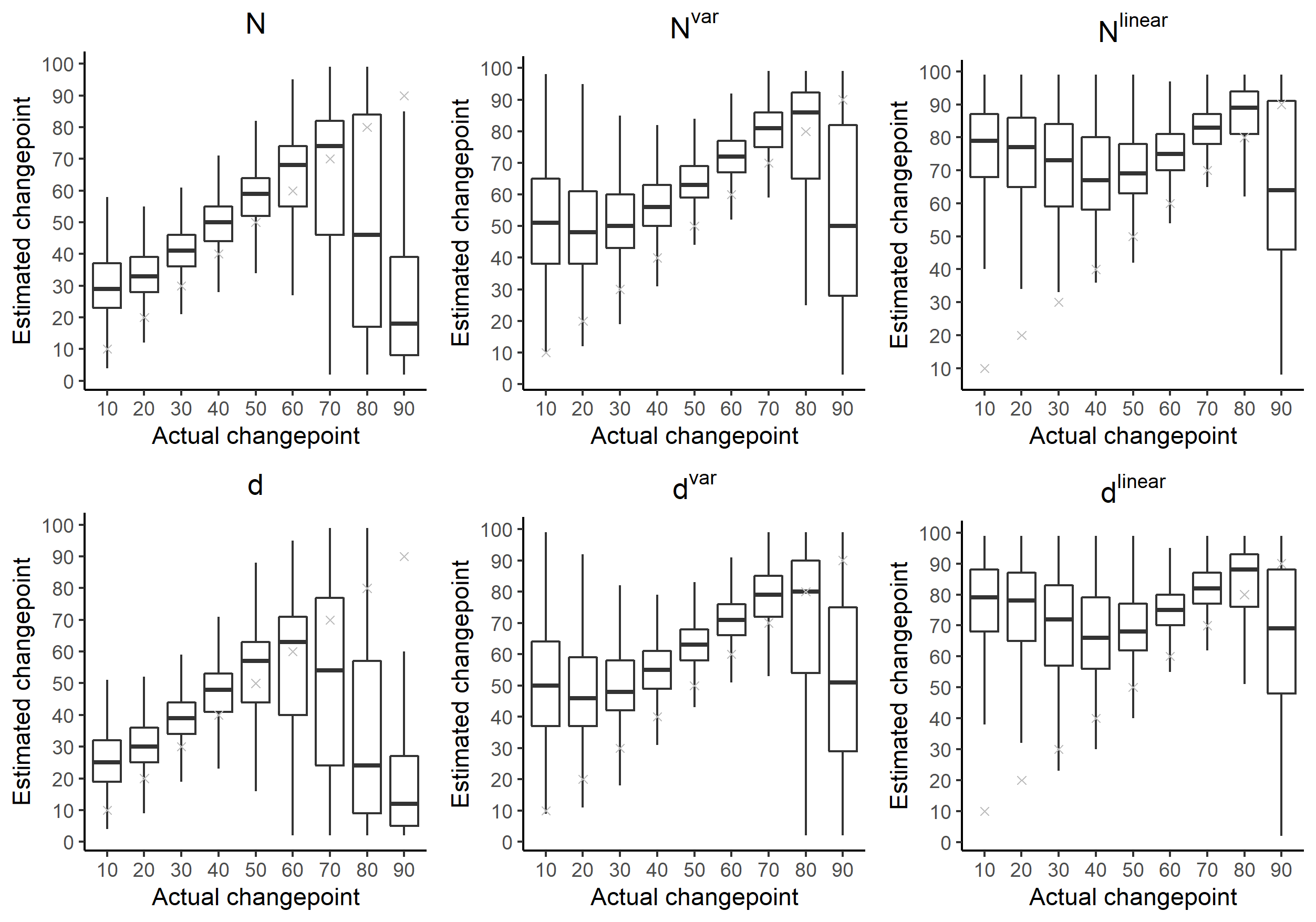}} \\
	\subfloat[$M=36$, $\theta = 0.05$]{\includegraphics[width=0.8\textwidth]{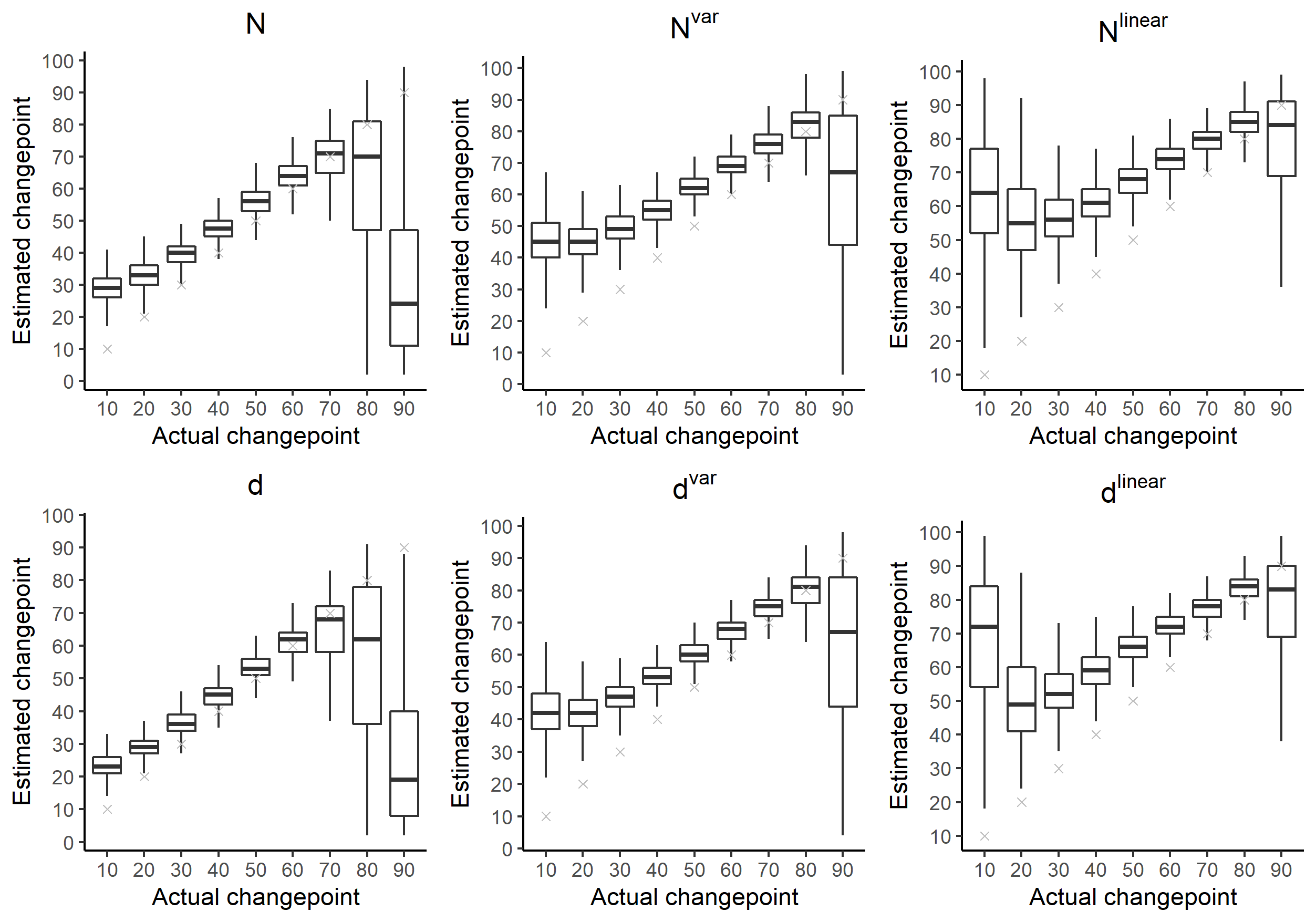}}\\ 
	\caption{Boxplots without outliers of the estimated changepoint versus the actual changepoint of $N$ and $d$-type statistics for Scenario~A ($T=100$). Crosses represent the actual changepoint \label{fig:estimatedrift_mean}}
\end{figure}

\begin{figure}
	\centering\captionsetup[subfloat]{labelfont=bf}
	\subfloat[$M=1$, $\theta = 0.10$]{\includegraphics[width=0.8\textwidth]{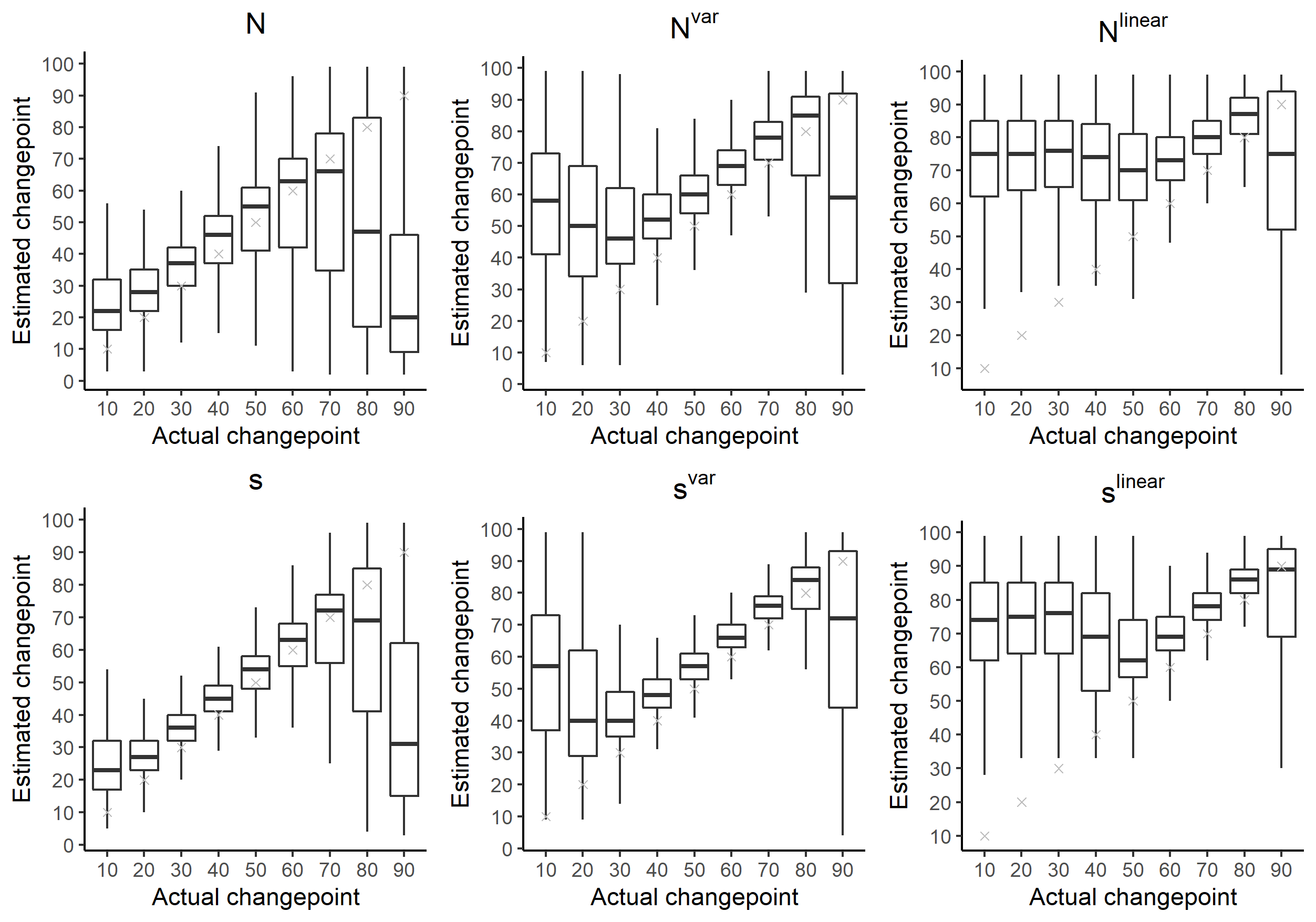}}\\
	\subfloat[$M=36$, $\theta = 0.05$]{\includegraphics[width=0.8\textwidth]{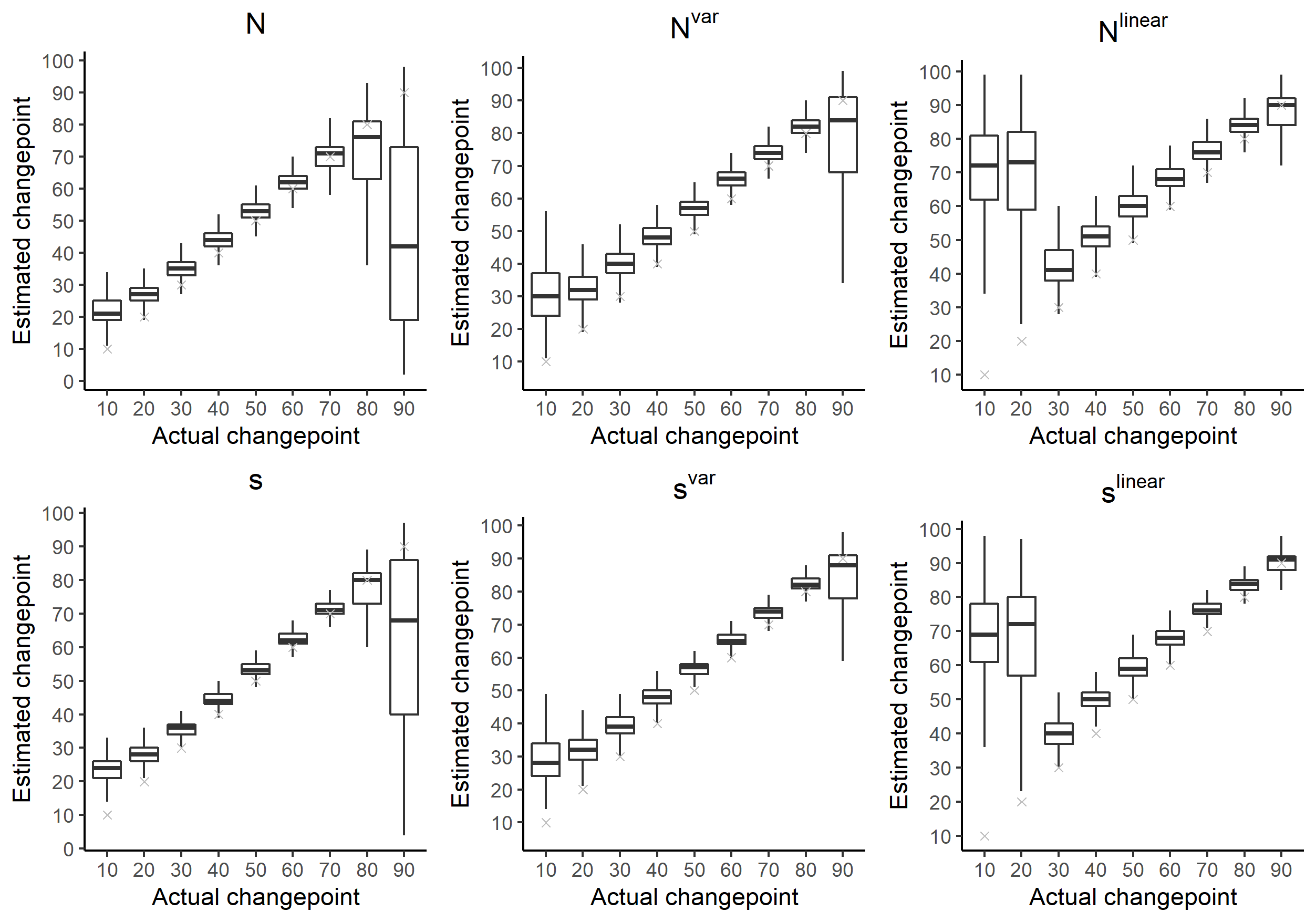}}\\ 
	\caption{Boxplots without outliers of the estimated changepoint versus the actual changepoint of $N$ and $s$-type statistics for Scenario~B ($T=100$). Crosses represent the actual changepoint \label{fig:estimatedrift_sd}}
\end{figure}

Figs.~\ref{fig:estimatedrift_mean}~and~\ref{fig:estimatedrift_sd} show boxplots of the estimated changepoint for the Scenarios A and B, respectively. We remark the following conclusions:
\begin{enumerate}
	\item As it was advanced in Section~\ref{sec:asymptotic}, the nonuniform variance in CUSUM-type statistics means that changepoints occurring near the data boundaries are more difficult to detect, hence, they have trouble in detecting changes occurring away from the middle of the series. This effect is reduced as the number of series $M$ or the magnitude of the drift $\theta$ increases.
	
	\item Comparing statistics with the same type of weight. In Scenario~A, $N$-type statistics place the changepoint slightly better than $d$-type. In Scenario~B, $s$-type statistics place the chagepoint better than $N$-type.
	
	\item The performance of the changepoints depends on the type of weight used. The statistics without weights properly place the changepoint when it is at the beginning or the middle of the series, but not at the end. The statistics with weights proportional to the inverse of the SD properly place the changepoint when it is not found at the beggining or the end of the series. The statistics with linear weights place the majority of changepoints in the second half of the series, so its estimate is not reliable for practical use, although this effect is reduced by increasing $M$.
\end{enumerate}

These changepoint detection tests based on the breaking of records only make use of the record occurrence to determine the changepoint estimate. For that reason, the estimated changepoint will usually be placed in the previous time of a record time, i.e., the effect of the drift is not immediately reflected in the observed record occurrence. This means that a proper estimate of the changepoint in the record occurrence will often be placed later in the series than the actual changepoint in the mean or variance. Thus, the main question here is then whether the correctly detected, but possibly displaced, changepoints are clustered near the actual value or not.

This section has been useful to illustrate the behavior of the tests against usual alternative hypotheses. Other scenarios could be considered, e.g., (C) a shift model in the mean, i.e., $\mu_t = \theta$ if $t_0 < t \leq T$ under Scenario~A; or (D) a mixture model with a drift in the right tail, i.e., $Y_{tm} = \epsilon^{(0)}_{tm}$ if $u_{tm} \leq \tau$ and $Y_{tm} = \mu_t + \epsilon^{(1)}_{tm}$ if $u_{tm} > \tau$ where $u_{tm} \sim U(0,1)$, $\tau$ a high quantile order (e.g., $\tau = 0.95$), $\mu_t$ under Scenario~A, and $\epsilon^{(0)}_{tm}$ and $\epsilon^{(1)}_{tm}$ truncate $N(0,1)$ in $(-\infty,\Phi^{-1}(\tau))$ and $(\Phi^{-1}(\tau),\infty)$, respectively. Preliminary analyzes show that the tests perform poorly against Scenario~C, but have great power against Scenario~D, even outperforming commonly used changepoint detection tests (e.g., the Pettitt test).

\section{Application to temperature series}
\label{sec:application}

To illustrate the practical use of the three types of records tests, we applied them to the daily maximum temperature series measured in degree Celsius ($^{\circ}$C) from $1940$ to $2019$ at Madrid, Spain. Data are provided by European Climate Assessment \& Dataset \citep[ECA\&D;][]{tank2002} available online at \url{https://www.ecad.eu}. Madrid is located in the center of the Iberian Peninsula ($40.4^{\circ}$ N, $3.7^{\circ}$ W) at $667$ meters a.s.l. and its daily temperature series has a seasonal component and a strong serial correlation. This series is analyzed using three different approaches to show the performance of the tests in different situations. The first approach considers the series of annual maximum temperatures, which corresponds to the traditional block maxima. The second approach considers the series of annual mean temperature. Finally, the series is considered on a daily scale. To do this, first we take $365$ subseries each corresponding to the data of a given day across years and then we select a subset of uncorrelated subseries \citep{cebrian2021} on which the procedure of Section~\ref{sec:seasonal} is applied. The three approaches have series of length $T = 80$, the first two with $M = 1$ and the third with $M = 58$ uncorrelated series out of the $365$ dependent subseries.

In the context of global warming, it is reasonable to assume an increasing trend in location that can cause an increase in the number of upper records as well as decrease the number of lower records with respect to the values expected under a stationary climate, i.e., IID series. For this reason, only results for $N$ and $d$-type statistics are shown. These statistics are powerful against this scenario and obtained more evidence than $s$-type statistics. To compare the detection time of a changepoint in location versus a changepoint in the record occurrence, we consider the \cite{pettitt1979} test, which is a nonparametric rank based test widely used to detect AMOC at location.

\begin{figure}
\centering\captionsetup[subfloat]{labelfont=bf}
	\subfloat[Maximum]{\includegraphics[width=0.48\textwidth]{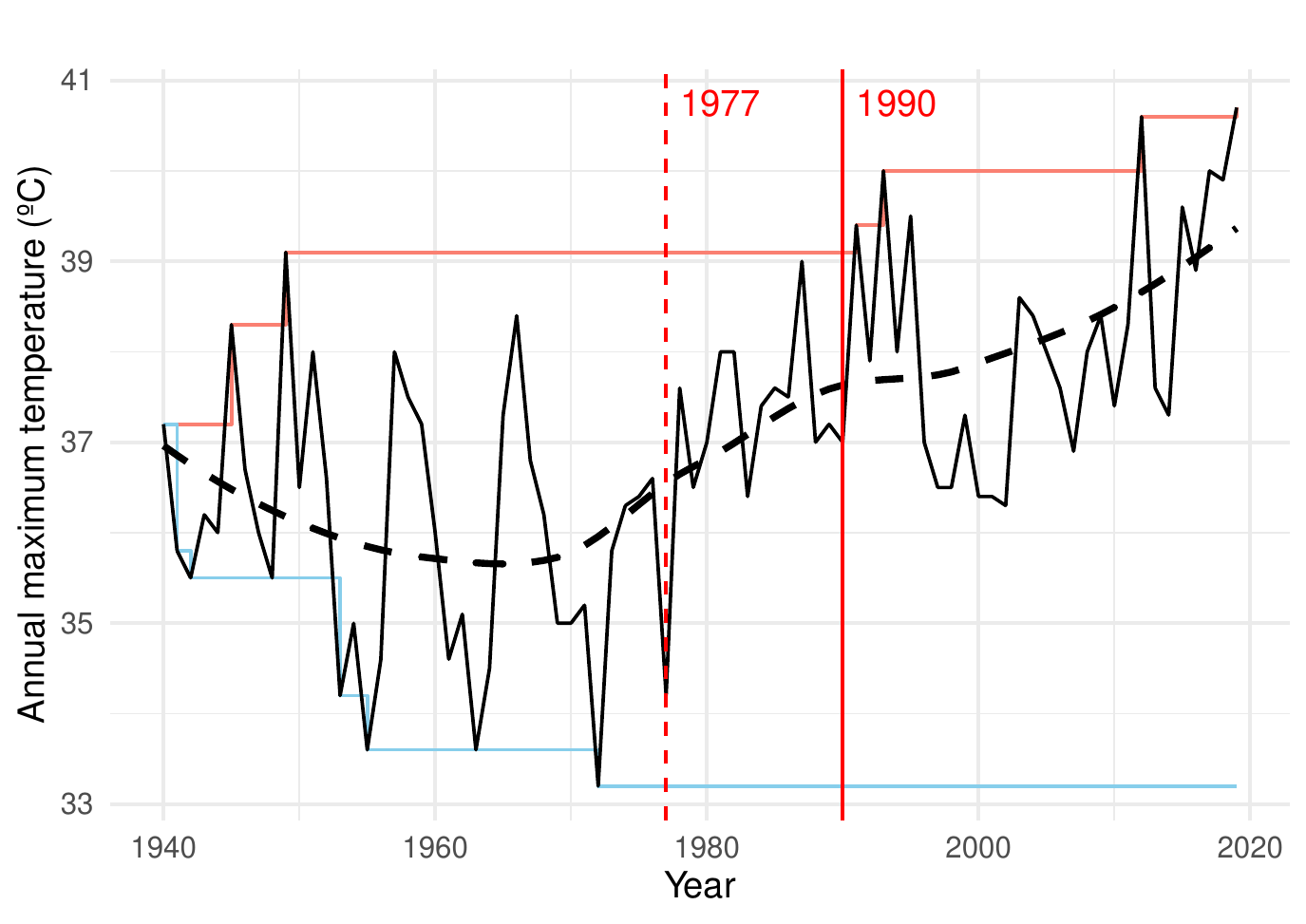}}
	\hfill
	\subfloat[Mean]{\includegraphics[width=0.48\textwidth]{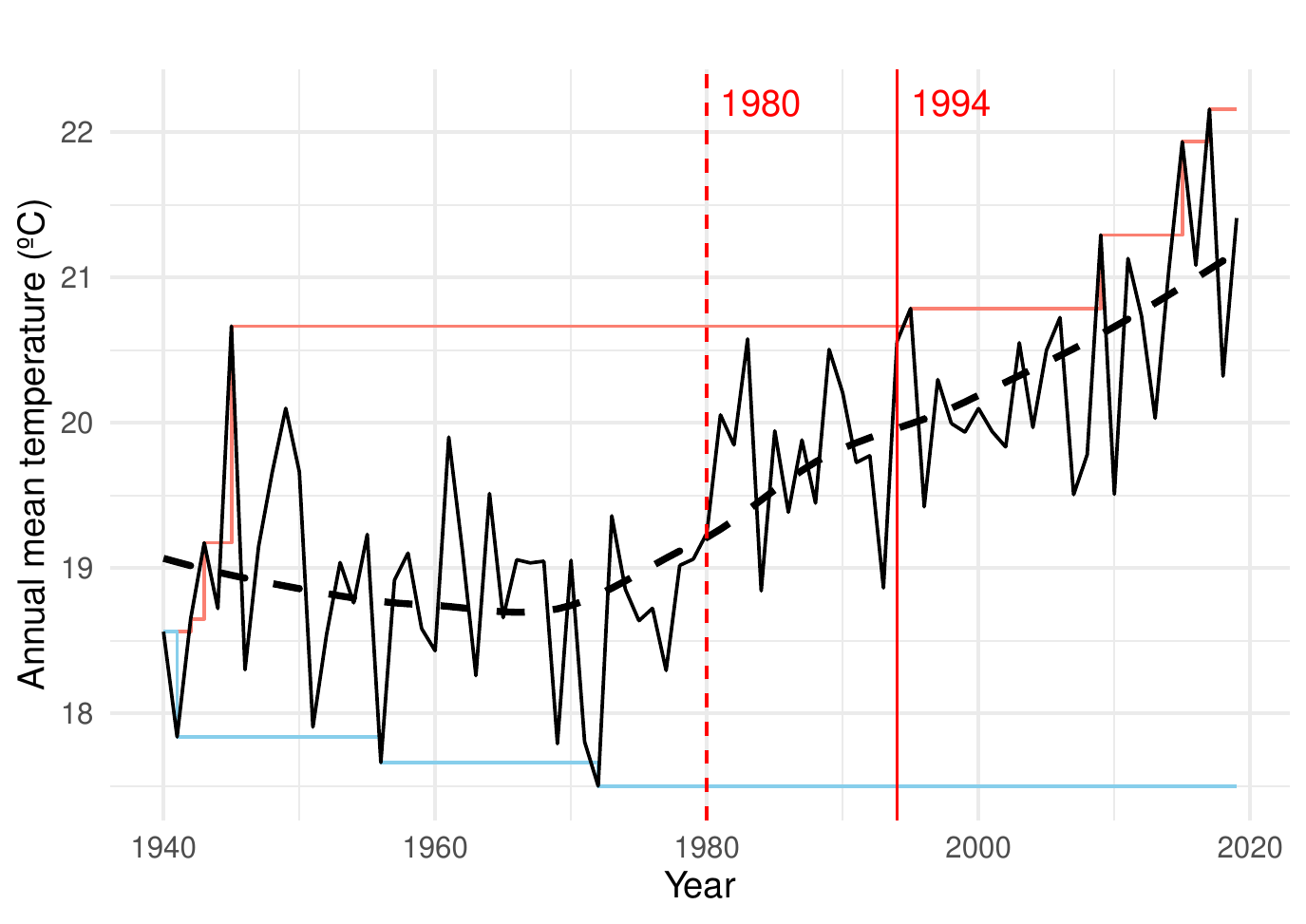}}\\ 
	\caption{Annual maximum \textbf{a} and mean \textbf{b} temperature series, and lower and upper records at Madrid, Spain. The vertical solid line is the estimated changepoint using the records tests, while the vertical dashed line is the estimated changepoint using the Pettitt test \label{fig:series}}
\end{figure}

Fig.~\ref{fig:series} shows time series plots of annual maximum \textbf{a} and annual mean \textbf{b} temperature at Madrid with their records and changepoint estimates. Table~\ref{tab:pvalues} shows for the two previous series and for the series in daily scale the p-values and changepoint estimates for the six records tests and the Pettitt test. Small p-values in the records tests provide evidence against the null hypothesis of stationarity, in particular, all tests are significant at a level $\alpha = 0.10$, all but one are significant at a level $\alpha = 0.05$ and fourteen out of eighteen are significant at $\alpha = 0.01$. The Pettitt test is also significant for any usual significance level in both series with $M = 1$. The estimated changepoint for the annual maximum temperature series is $\hat{t}_0 = 51$ (year $1990$) for all the records statistics and $\hat{t}_0 = 38$ ($1977$) for the Pettitt test. The minimum p-value for the records tests is $0.0013$ for the statistic $N^{var}$. The estimated changepoint for the annual mean temperature series is $\hat{t}_0 = 55$ ($1994$) with the statistics without weights and with weights proportional to the inverse of the SD, but it is $\hat{t}_0 = 69$ ($2008$) for the statistics with linear weights and $\hat{t}_0 = 41$ ($1980$) for the Pettitt test. The minimum p-value of the records tests is $0.0004$ for the tests $N^{var}$ and $N^{linear}$. For the daily scale series the changepoint estimate is $\hat{t}_0 = 38$ ($1977$) for all records statistics and here the minimum p-value is $4e-05$ for $d^{var}$.

The results in Table~\ref{tab:pvalues} agree with the results obtained in Section~\ref{sec:simulation}. When $M = 1$, $N$-type statistics obtain lower p-values than $d$-type, and the statistics with weights proportional to the inverse of the SD are those that obtained the strongest evidence. The changepoint estimate of the records tests is usually placed between 10 and 15 years after the Pettitt test estimates a changepoint in location. The changepoint estimated by the statistics with linear weights tend to locate the change very late. It is noteworthy that the changepoint is always estimated just before a record (see Fig.~\ref{fig:series}), so the changepoint estimate of a significant records test can be interpreted as the time from which there is evidence that the record occurrence is no longer stationary and the tail of the distribution begins to take on ever greater values, not previously seen. When $M > 1$ the results are more stable, the estimated changepoint appears earlier as more information is available and $d$-type statistics obtain smaller p-values than $N$-type.

\begin{table}\centering
	\caption{Estimated changepoint $\hat{t}_0$ and p-value of the records tests and the Pettit test for the annual maximum, annual mean and daily scale temperature series at Madrid, Spain. The weighted statistics used 1,000,000 replicates to estimate the p-value \label{tab:pvalues}}
	\begin{tabular}{lcccccccc}
		\hline\noalign{\smallskip}
		Statistic & \multicolumn{8}{c}{Series} \\
		\noalign{\smallskip}\hline\noalign{\smallskip}
		& \multicolumn{2}{c}{Annual maximum} & \phantom{abc}& \multicolumn{2}{c}{Annual mean} &
		\phantom{abc} & \multicolumn{2}{c}{Daily}\\
		& $\hat{t}_0$ & p-value & & $\hat{t}_0$ & p-value & & $\hat{t}_0$ & p-value \\ 
		\noalign{\smallskip}\hline\noalign{\smallskip}
		$N$          & $51$ & $0.0031$ & & $55$ & $0.0030$ & & $38$ & $0.0003$ \\
		$N^{var}$    & $51$ & $0.0013$ & & $55$ & $0.0004$ & & $38$ & $0.0002$ \\
		$N^{linear}$ & $51$ & $0.0090$ & & $69$ & $0.0004$ & & $38$ & $0.0328$ \\
		$d$          & $51$ & $0.0443$ & & $55$ & $0.0728$ & & $38$ & $0.0013$ \\
		$d^{var}$    & $51$ & $0.0016$ & & $55$ & $0.0015$ & & $38$ & $4e-05$  \\
		$d^{linear}$ & $51$ & $0.0115$ & & $69$ & $0.0018$ & & $38$ & $0.0029$ \\
		Pettitt      & $38$ & $9e-07$ & & $41$ & $8e-10$ & & -  & - \\
		\noalign{\smallskip}\hline
	\end{tabular}
\end{table}

\begin{figure}
	\centering\captionsetup[subfloat]{labelfont=bf}
	\subfloat[Annual maximum]{\includegraphics[width=0.48\textwidth]{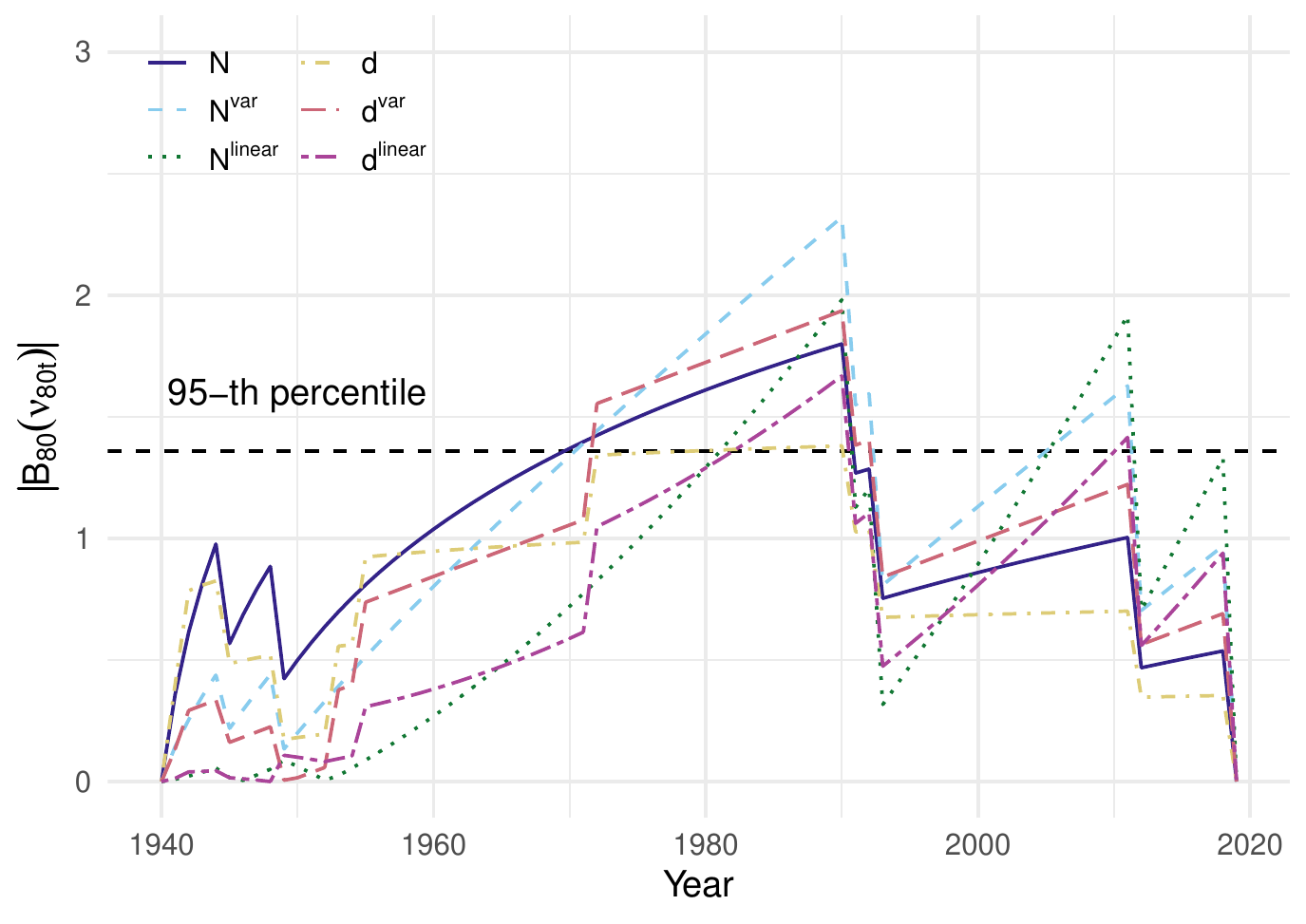}}
	\hfill
	\subfloat[Annual mean]{\includegraphics[width=0.48\textwidth]{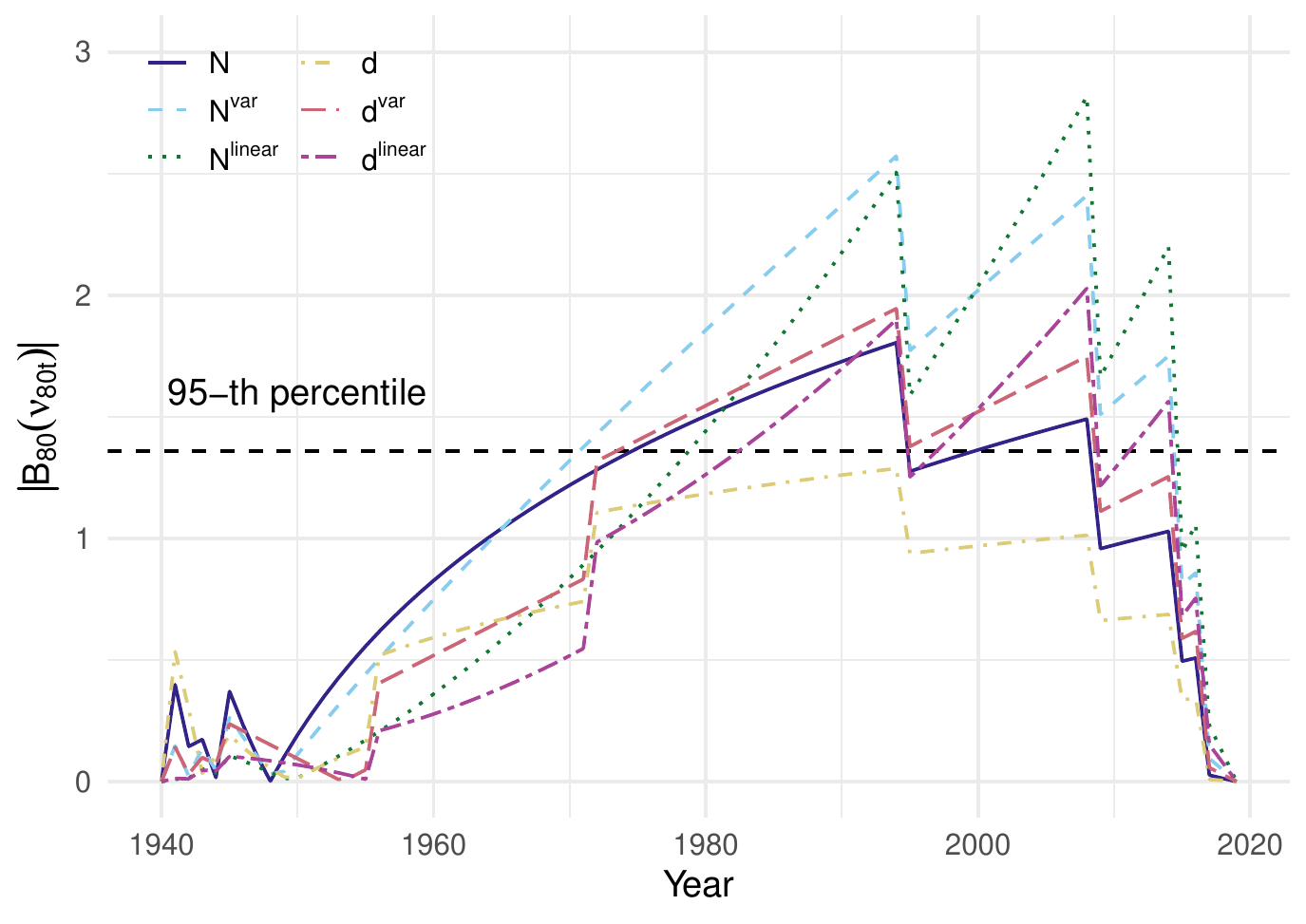}}\\
	\subfloat[Daily]{\includegraphics[width=0.48\textwidth]{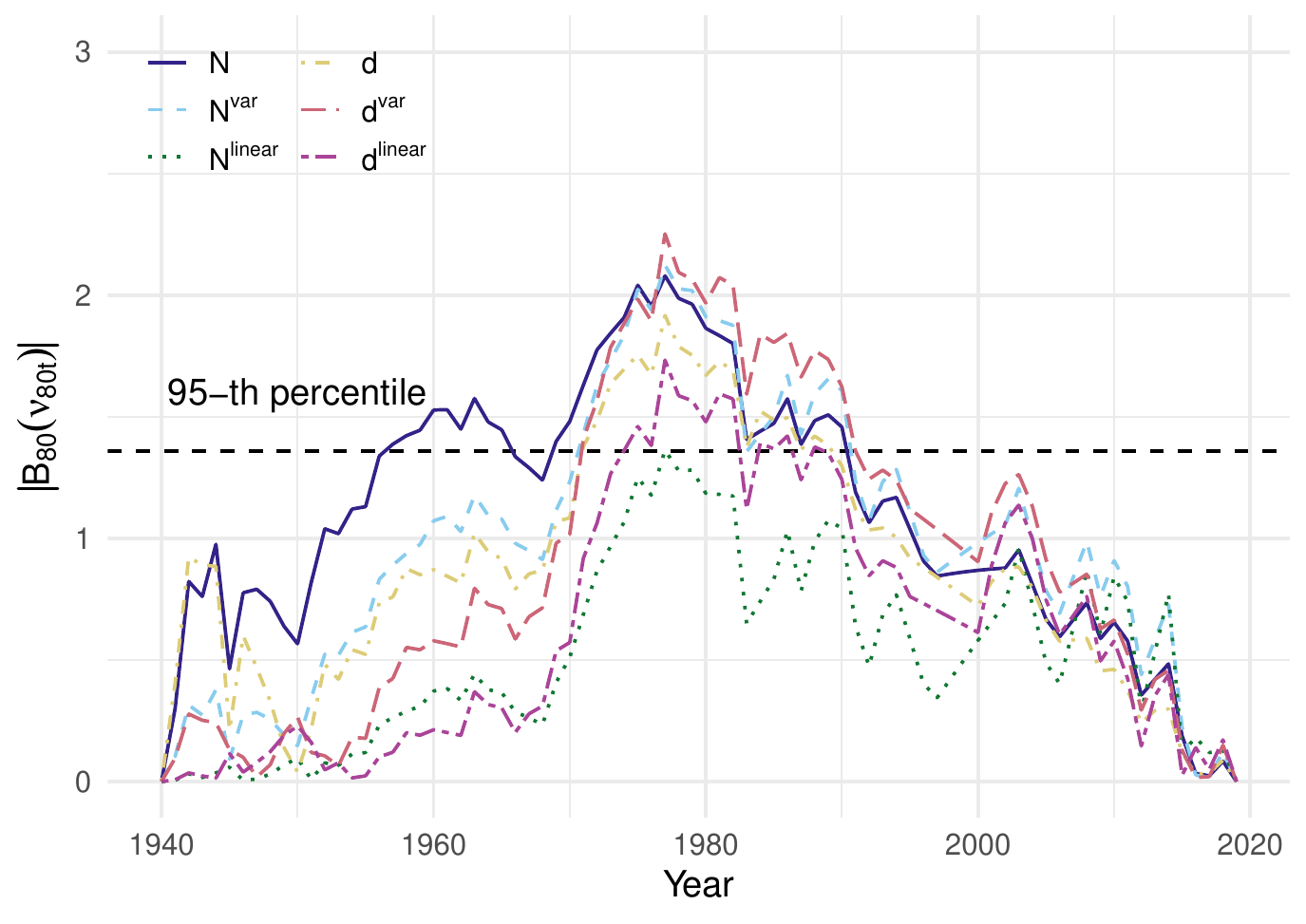}} 
	\caption{Absolute value of the processes associated with the records statistics for changepoints of annual maximum \textbf{a}, annual mean \textbf{b} and daily \textbf{c} temperature series at Madrid, Spain. The horizontal dashed line represents the $95$-th percentile of the Kolmogorov distribution \label{fig:statistic1}}
\end{figure}

Fig.~\ref{fig:statistic1} plots the year versus the absolute value of the processes associated with the records statistics for the annual maximum \textbf{a} and mean \textbf{b} temperature series along with $95\%$ confidence thresholds based on the Kolmogorov distribution (they are very similar even for nonKolmogorov distributed statistics). These plots allow to see the evolution of the processes and other possible points with greater record probability than under the null hypothesis. Again, the stationary null hypothesis is rejected, indicating potential changepoint $1990$ and $1994$, respectively. The equivalent plot for the daily scale temperature series is shown in \textbf{c}, showing a clear maximum in $1977$.

\section{Discussion, conclusions and future work}
\label{sec:conc}

The interest in statistical tools to analyze nonstationary behaviors in the extreme values of the distribution is growing. While extreme value analysis has been traditionally based on block maxima and excesses over threshold, this paper proposes the use of records to study changes in the tails of the distribution. In particular, this paper proposes three novel distribution-free changepoint detection tests and some generalizations based on the breaking of records to (1)~detect changes in the extreme events of the distribution, (2)~learn about features of the record occurrence and (3)~analyze data when only their records are available.

The proposed statistics are CUSUM-type statistics based on the record indicators. Statistics to deal with seasonal series have also been considered. Despite having a very small sample information compared to the total of the series, the Monte Carlo simulations have shown that the proposed records tests are capable of detecting deviations from the null hypothesis and a reasonable changepoint at which this deviation becomes significant in the probabilities of record. However, care must be taken in the interpretation of the changepoint estimate, on the one hand it is usually misplaced when the actual changepoint is located in the ends of the series. On the other hand, when it is well located, it is often slightly after the actual changepoint in location or scale if it exists, i.e., the effect of the change is not immediately reflected in the observed record occurrence. 

The recommendation for use according to the power and changepoint estimate accuracy  of the tests is as follows. If an increase in the number of records with respect to the stationary case is expected in a single tail of the distribution, the results show that $N$-type statistics are usually recommended. If an increase in the number of records is expected in both tails, then $s$-type statistics are preferred. The statistics without weights have the advantage of having a known asymptotic distribution, while the statistics with weights proportional to the inverse of the SD have shown to have a more balanced behavior against the alternative hypothesis in the simulation results, with the disadvantage that their distribution must be calculated using Monte Carlo techniques.

The proposed tests join two important aspects in the study of climate change, changepoint detection methods and record-breaking events. This last concern has been made apparent when applying the tests on different summary series (block maxima and annual mean) and on the series on a daily scale of temperatures at Madrid, Spain; detecting significant evidence of warming since the late 70s and early 90s. 

Future work may go in different directions. (1)~Combining the information from the different statistics could be of interest to increase the power and decrease the chance of mis-detection, e.g., the harmonic mean p-value by \cite{wilson2019} could be used to have a single p-value of all tests. (2)~The idea of splitting the series is fundamental to dealing with seasonal behavior. Here we use the method by \cite{cebrian2021} to extract uncorrelated subseries. Another alternative to consider would be to implement permutation tests, i.e., the test statistic under the null hypothesis would be obtained by calculating all possible values of the test statistic under all possible rearrangements of the observed years, $t=1,\ldots,T$. In this way we maintain the dependence structure between the subseries without the need to have a subset of independent subseries. (3)~Our method has been developed within the AMOC domain, however its extension to the multiple changepoint domain could be of interest. The simplest procedure would be to split the series where the changepoint is detected and retest the two subseries separately. However, this can cause the number of records in the new subseries to be too small to detect new changepoints, so other alternatives should be studied.
 
Finally, it is noteworthy that the proposed changepoint detection records tests are not only useful for analyzing the effect of global warming on the occurrence of records, but also in other fields where records are important. Other applications of these tests are in other environmental sciences in the presence of climate change, in the study of extreme values in stock prices or in the influence that new sports equipment has on the occurrence of sports records. To facilitate its use, all the statistical tools proposed in this paper are included in the R package \texttt{RecordTest} \citep{recordtest} available from CRAN at \url{https://CRAN.R-project.org/package=RecordTest}.

\begin{acknowledgements}
This work was partially supported by the Ministerio de Ciencia e Innovaci\'on under Grant PID2020-116873GB-I00; Gobierno de Arag\'on under Research Group E46\_20R: Modelos Estoc\'asticos; and Gobierno de Arag\'on under Doctoral Scholarship ORDEN CUS/581/2020. The author thanks Jes\'us As\'in and Ana C. Cebri\'an for helpful comments; the editor and two anonimous reviewers for helpful reviews; and the ECA\&D project for providing the data.

This version of the article has been accepted for publication at \emph{Environmental and Ecological Statistics}, after peer review but is not the Version of Record and does not reflect post-acceptance improvements, or any corrections. The Version of Record is available online at: \url{https://doi.org/10.1007/s10651-022-00539-2}.
\end{acknowledgements}

\noindent\textbf{Data availability} Data and metadata are provided by the ECA\&D project and available at \url{http://www.ecad.eu}. The series of Madrid is the blended series of station SPAIN, MADRID - RETIRO (STAID: 230).

%
\section*{Declarations}

\noindent\textbf{Conflict of interest} The author declares that he has no conflict of interest.

\noindent\textbf{Ethical approval} This work does not contain any studies with human participants and/or animals.

\appendix

\section{Appendix: The variance of $B_T(\nu_{Tt})$} \label{app:var}

\begin{figure}
	\begin{center}
		\includegraphics[width=\textwidth]{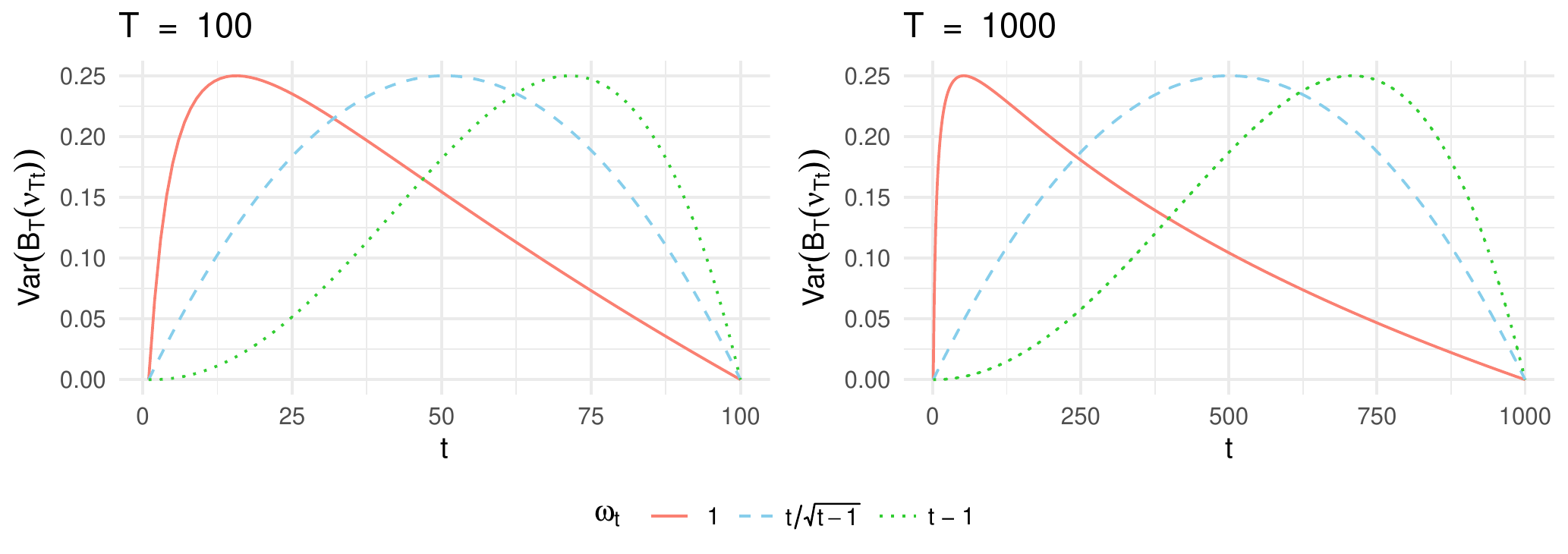}
	\end{center}
	\caption{The $Var(B_T(\nu_{Tt})) = \nu_{Tt}(1 - \nu_{Tt})$ across $t=1,\ldots,T$ ($T = 100, 1000$) and weights $\omega_t$ given in Section~\ref{sec:w}.  \label{fig:VARB}}
\end{figure}

Fig.~\ref{fig:VARB} shows the variance of $B_T(\nu_{Tt})$ for $T = 100$ and $1000$ across $t=1,\ldots,T$.  In particular, it is shown for the unweighted statistic, the statistics with weights proportional to the inverse of the SD and linear weights (see Section~\ref{sec:w}). While the second one generates a symmetric variance in $\{1,\ldots,T\}$ by construction, this does not happen with the other two. In all three cases the variance is zero for $t \in \{1,T\}$ and the maximum value is $1 / 4$. 

The nonuniform variance makes changepoints occurring near the beginning or the end of the series (small variance times) more difficult to detect. Under the null hypothesis, the process reaches its maximum (in absolute value) with the highest probability at time $t$ where it has the highest variance. Then, deviations from the null hypothesis at small variance times generate smaller deviations in $B_T(\nu_{Tt})$ than deviations at times of maximum variance. Thus, it is expected that the unweighted statistic will have more power when the changepoint is at the beginning of the series, the statistic with weights proportional to the inverse of the SD when the changepoint is in the middle of the series and the statistic with linear weights when the changepoint is at the end of the series. These conclusions agree with the analysis of power in Section~\ref{sec:power}.

\section{Appendix: Proof of Proposition 2.1} \label{app:proof}

We prove that the weighted statistics with polynomial weights $\omega_t \sim t^n$ as $t \rightarrow \infty$ for $n > 0$ do not have asymptotic Gaussian properties. In particular, the distribution of the weighted number of records do not approach the normal distribution for increasing $T$. This is verified by using its asymptotic skewness and showing that it is different from $0$ (the skewness of any normal RV) for $n > 0$. As a consequence, the asymptotic distribution of the functional evolution of the weighted number of records does not approach that of the Wiener process.

\begin{proof}[of Proposition~\ref{prop:normal}]
	To prove that the $\xi_{Tt}^\omega$'s do not satisfy the central limit theorem, it is sufficient to prove that the sum $N_T^{\omega} = \sum_{t=1}^{T} \omega_t I_t$ does not have skewness $0$ as $T \rightarrow \infty$. Using the basic properties of the central moments of a RV,
	\begin{gather*}
	\mu \equiv \mu_1(N_T^{\omega}) = E(N_T^{\omega}) = 
	\sum_{t=1}^{T} \omega_t \, \frac{1}{t},\\
	\sigma^2 \equiv \mu_2(N_T^{\omega}) = E\left[(N_T^{\omega} - \mu)^2\right] =
	\sum_{t=1}^{T} \omega_t^2 \, \frac{t-1}{t^2},\\
	\mu_3(N_T^{\omega}) = E\left[(N_T^{\omega} - \mu)^3\right] = 
	\sum_{t=1}^{T} \omega_t^3 \, \frac{t^2 - 3t + 2}{t^3}.
	%
	\end{gather*}
	Then, the following is a consequence of the properties of the generalized harmonic numbers, as $T \rightarrow \infty$,
	\begin{equation*}
	Skew(N_T^{\omega}) = \frac{\mu_3(N_T^{\omega})}{\sigma^3} \sim \frac{\sum_{t=1}^T t^{3n-1}}{\left(\sum_{t=1}^T t^{2n-1}\right)^{3/2}} \longrightarrow \frac{2}{3} \sqrt{2n}.
	\end{equation*}
	Consequently the skewness of $N_T^{\omega}$ is asymptotically different from $0$ for $n > 0$. \qed
\end{proof} 

This proposition is easily extended to the weighted statistics based on the $d_t$'s and $s_t$'s. The former requires the calculation of the kurtosis since its skewness is $0$ because it is a symmetric RV. We omit the details for the sake of brevity.

\bibliographystyle{spbasic}      
\bibliography{arxiv_v2}   

\end{document}